# Growth Techniques for Bulk ZnO and Related Compounds

Detlef Klimm, Detlev Schulz, Steffen Ganschow, Zbigniew Galazka, Reinhard Uecker

Leibniz Institute for Crystal Growth, Max-Born-Str. 2, 12489 Berlin, Germany

## ABSTRACT

ZnO bulk crystals can be grown by several methods. 1) From the gas phase, usually by chemical vapor transport. Such CVT crystals may have high chemical purity, as the growth is performed without contact to foreign material. The crystallographic quality is often very high (free growth). 2) From melt fluxes such as alkaline hydroxides or other oxides ($MoO_3$, $V_2O_5$, $P_2O_5$, PbO) and salts ($PbCl_2$, $PbF_2$). Melt fluxes offer the possibility to grow bulk ZnO under mild conditions (<1000°C, atmospheric pressure), but the crystals always contain traces of solvent. The limited purity is a severe drawback, especially for electronic applications. 3) From hydrothermal fluxes, usually alkaline (KOH, LiOH) aqueous solutions beyond the critical point. Due to the amphoteric character of ZnO, the supercritical bases can dissolve it up to several per cent of mass. The technical requirements for this growth technology are generally hard, but this did not hinder its development as the basic technique for the growth of α-quartz, and meanwhile also of zinc oxide, during the last decades. 4) From pure melts, which is the preferred technology for numerous substances applied whenever possible, e.g. for the growth of silicon, gallium arsenide, sapphire, YAG. The benefits of melt growth are (i) the high growth rate and (ii) the absence of solvent related impurities. In the case of ZnO, however, it is difficult to find container materials that are compatible from the thermal (fusion point $T_f$ = 1975°C) and chemical (required oxygen partial pressure) point of view. Either cold crucible (skull melting) or Bridgman (with reactive atmosphere) techniques were shown to overcome the problems that are inherent to melt growth. Reactive atmospheres allow to grow not only bulk ZnO single crystals, but also other TCOs such as $\beta\text{-}Ga_2O_3$ and $In_2O_3$.

## INTRODUCTION

The semiconducting properties of zinc oxide have been used for years for the production of voltage dependent resistors (varistors). In such devices *p-n* junctions are formed in the grain boundaries of ceramic bodies. Moreover, doped ZnO can be used as TCO (transparent conducting oxide) e.g. for transparent electrodes. Still the production of single crystalline heteropolar devices, especially LED's, is prohibited by the fact that efficient *p*-type doping (high mobility and sufficient hole density) was not yet possible. Further work towards ZnO based devices requires not only improved epitaxy techniques, but also the delivery of substrates for homoepitaxy with good quality.

Principally, ZnO single crystals and substrates made thereof are available nowadays. The huge variety of growth techniques that can be used for the production of bulk ZnO is surprising, and perhaps no other substance is really grown by so many different methods. The benefits and drawbacks of these methods will be reviewed in this paper, and a short outlook on the bulk growth of alternative TCO materials will be given.

## GROWTH TECHNIQUES FOR ZnO SORTED BY PHASE TRANSITION

The chemical substance zinc oxide is a white powder, which is only apparently chemically stable. Instead, it decomposes easily to its components (1), or reacts with other substances such as carbon dioxide (2) or water (3):

$$Zn + \frac{1}{2}O_2 \leftrightarrow ZnO \quad \Delta H = -350.5 \text{ kJ/mol} \tag{1}$$

$$ZnO + CO_2 \leftrightarrow ZnCO_3 \quad \Delta H = -68.6 \text{ kJ/mol} \tag{2}$$

$$ZnO + H_2O \leftrightarrow Zn(OH)_2 \quad \Delta H = +2.4 \text{ kJ/mol} \tag{3}$$

(Enthalpies are given for normal conditions). Carbon dioxide and water vapor are with 390 ppm or ≈10,000 ppm, respectively, constituents of ambient air — hence the surface of ZnO powder as well as ZnO crystals is usually covered by layers of zinc carbonate and/or hydroxide (Fig. 1).

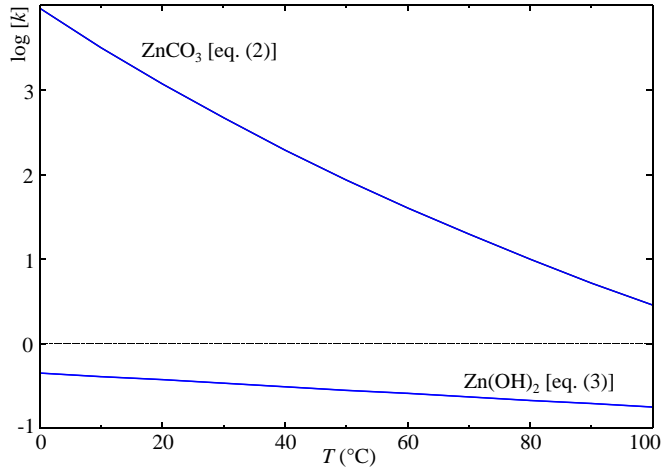

**Fig. 1: Equilibrium constants $k(T)$ for the chemical reactions (2) and (3). It is obvious that ZnO if stored under ambient air is always contaminated by carbonate and to a smaller degree by hydroxide.**

Zinc hydroxide (3) is amphoteric. This means it can easily be dissolved by weak aqueous acids (4) and by strong bases (5):

$$Zn(OH)_2(aq) + 2H^+(aq) \leftrightarrow Zn^{2+}(aq) + 2H_2O \quad \Delta H = -203.8 \text{ kJ/mol} \tag{4}$$

$$Zn(OH)_2(aq) + OH^-(aq) \leftrightarrow [Zn(OH)O]^-(aq) + H_2O \quad \Delta H = +8.7 \text{ kJ/mol} \tag{5}$$

Not only hydroxide zincates such as $[Zn(OH)_4]^{2-}$ or $[Zn(OH)O]^-$ (5) can be formed in solution, but also other complex ions such as ammono zincates e.g. $[Zn(NH_3)_4]^{2+}$.

The high reactivity of zinc metal and zinc oxide, which is expressed by the equilibrium reactions (1) – (5), allows the application of a big variety of growth techniques for ZnO bulk crystals. Nearly all technical processes that were developed for growing bulk single crystals can be applied for this material. In the next subsections, these techniques for ZnO bulk crystal growth will be categorized according to the phase transformation that enables the crystallization of ZnO from a fluid mother phase.

Besides it should be mentioned that diethylzinc $(C_2H_5)_2Zn$ is the first organometallic compound that was ever prepared (in 1849 by Frankland from Zn metal and ethyl iodide [1]). Compounds of this type are typically used for the MOCVD (metal organic chemical vapor deposition) of thin layers, but were so far not used for bulk crystal growth.

**From the gas phase**

The formation of ZnO from Zn metal and oxygen is strongly exothermal (1): Liquid zinc burns in air under the formation of white ZnO fume. On an industrial scale, Zn metal can be evaporated and reacts with air in the indirect (French) process to ZnO. Upon heating, ZnO dissociates completely and the vapor pressure of its constituents Zn and $O_2$ becomes so large, that the solid starts to evaporate. The evaporation rate depends on the surrounding atmosphere: in thermogravimetry measurements the mass loss becomes remarkable at >1200°C in vacuum ($10^{-6}$ bar), >1300°C in argon, and >1450°C in air [2, Fig. 7 there]. For $T > 1600$°C the vapor pressure becomes so large that reasonable transport rates for sublimation growth (physical vapor transport, PVT) can be obtained [3]. It should be noted, however, that such crystals are typically only on the millimeter scale.

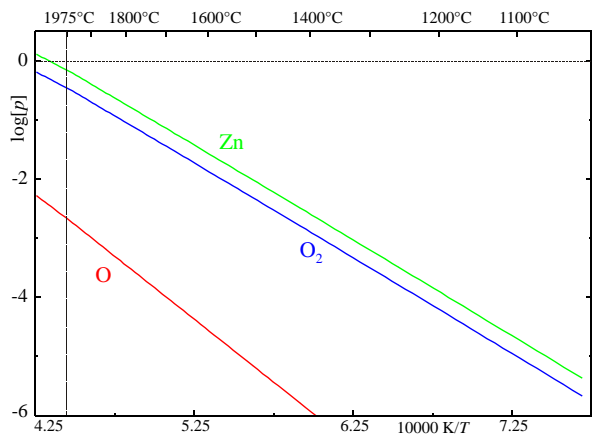

**Fig. 2: Vapor pressure of different species over ZnO. 1975°C is the melting point of ZnO. The horizontal line corresponds to 1 bar.**

Significantly better crystals can be grown by chemically assisted vapor phase transport (CVT). In Erlangen (Germany) the growth of bulk ZnO by different techniques was performed since the 1930s (see next subsection), and there E. Scharowsky [4] performed a PhD work based on an apparatus that is schematically drawn in Fig. 3. There liquid Zn metal (melting point 419°C) was held inside a ceramic tube at ca. 600°C. It should be noted that the volatility of Zn is way higher compared to ZnO, and reaches at 600°C already 15 mbar (boiling point at ambient pressure is ca. 1200°C). A flow of ca. 1100 ml/min oxygen-free nitrogen (with ca. 60 ml/min hydrogen added to avoid accidental Zn oxidation) transports the evaporated zinc via a conical tube outlet to a second, larger furnace that is held at higher temperature (maximum 1200°C). This furnace is partially open and allows the inflow of air. The $N_2$ flux must be adjusted in such a manner that backflow of air to the Zn metal is prohibited, but on the other side a too strong "blow" prohibits crystal growth at the rim of the cone. Under optimum conditions colorless ZnO needles with up to 4 cm length and a few tenth of a millimeter diameter with almost perfect hexagonal cross section could be obtained. Sometimes lancet shaped single crystals were harvested. A too large gas flow results in microcrystalline ZnO fume, under a too large $H_2$ flux the crystals become yellowish.

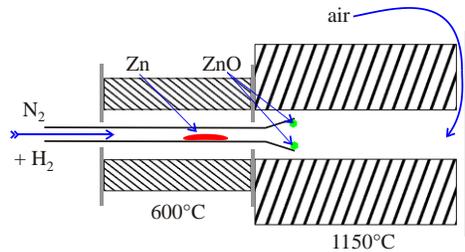

**Fig. 3: The setup that was used by Scharowsky (1953) [4] for the CVT growth of ZnO single crystals.**

Later this technique was modified by Helbig [5] who used annealed ZnO feed as source material. With similar $N_2/H_2$ flow rates through the feedstock, and a separately controlled flow of pure $O_2$ to the growth zone, ZnO crystals up to 20 g mass were grown (diameter up to 7 mm). One can assume that the considerably higher and well controlled growth temperature (1600±0.2 K instead of 1423 K) enabled the larger and more isotropic growth rate, compared to Scharowsky [4]. The material that was offered by Eagle-Picher, Inc. is produced in a similar process based on ZnO transport by $H_2$ [6].

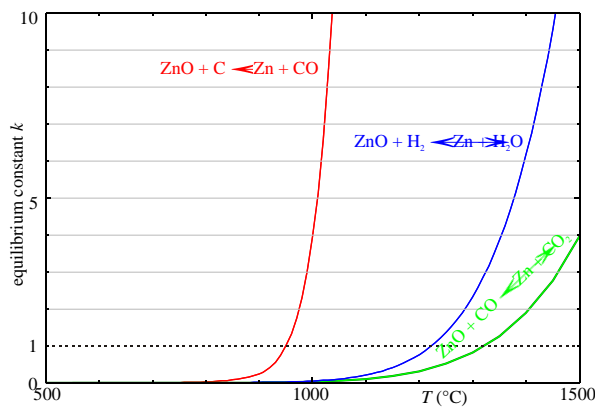

**Fig. 4: Equilibrium constants for different chemical reactions enabling CVT growth of ZnO (total pressure = 1 bar).**

Chemical vapor transport relies on an equilibrium reaction between the substance to be grown and some transport agent. If $H_2$ is used, like in the examples described so far, this is the reaction that labels the middle (blue) curve in Fig. 4. Reasonable growth rates require conditions yielding equilibrium constant $k \approx 1$. This is the case around 1200°C.

Admixtures of solid carbon to the feedstock (source temperature around 1000°C [7]) or carbon monoxide gas flow (source temperature up to 1300°C [8]) are alternative transport agents for CVT growth of ZnO and are depicted by the red and green curves in Fig. 4, respectively.

It is an advantage of all growth processes from the gas phase that the crystals grow mainly free from mechanical loads (except minor thermal stresses) and in a relatively clean environment where besides Zn and O only the transport agent is present. Consequently, the crystals often show superior quality by means of chemical purity as well as crystallographic quality (e.g. rocking curve width for 2Θ scans FWHM < 1 arcmin, carrier mobility >180 cm$^2$/Vs). Two aspects should be mentioned that might be less beneficial: In the case of carbon, C can be incorporated if the oxygen partial pressure in the system is too low [9], and hydrogen or hydrogen related point defects are known to play a dominant role in the electrical properties on ZnO [10].

**From the melt**

Already 20 years before Scharowsky's [4] vapor growth experiments in the same laboratories in Erlangen annealing experiments were performed. Fritsch (1935) [11] pressed ZnO powder to cylinders with length up to 20 mm and diameter up to 9 mm (Fig. 5a). A resistance heater was used to heat up these cylinders to 800°C and reduce the electrical resistivity of the semiconducting material. In a second step, AC current up to 15 A was passed through the cylinders for several hours, resulting in a pyrometrically measured surface temperature up to 1500°C. These experiments were performed in different atmospheres under different total pressure, up to 120 bar of pure oxygen. Sometimes clear drops of molten and solidified ZnO appeared — obviously the first molten ZnO ever produced! Experiments to melt pure ZnO in an iridium container failed because Ir was oxidized. Similar experiments with *rf* heating were performed later by Burmeister [12].

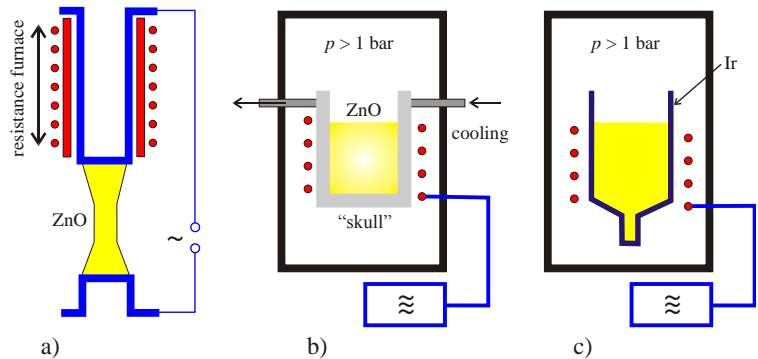

**Fig. 5: 3 different methods that were developed for the growth of bulk ZnO from the melt: a) Direct heating of ZnO ceramic by AC current [11]. b) Skull melting by *rf* energy coupling directly into ZnO melt [14]. c) Bridgman growth with iridium crucible and reactive atmosphere [15].**

Skull melting is an alternative method for melt growth which was developed for ZnO by Nause and is commercialized by CERMET Inc. [13,14]. Here *rf* heating energy is coupled directly into the ZnO melt that is surrounded by a cooled "skull" of solid ZnO powder (Fig. 5b). ZnO crystals grown by this method are chemically very pure because the material has no contact to any foreign substance. The atmosphere can be controlled in a very wide range, which is beneficial for the Zn/O stoichiometry of the grown crystals. Nevertheless, skull melting has one severe drawback: The electrical conductivity of dielectric melts rises usually with temperature. Consequently, "hot spots" inside the melt volume are heated more than colder volume elements, and thermal gradients are boosted.

Another concept for ZnO melt growth relies on "reactive atmospheres" with self-adjusting oxygen partial pressure $p_{O_2}(T)$ and reduces problems that can occur if iridium crucible metal is used in an oxidizing atmosphere [15,16]. A conflict arises from the dissociation reaction (1) that leads to a decomposition of ZnO at high *T* which could be reduced (according the law of mass action) by a high $p_{O_2}$, and the oxidation reaction

$$\mathrm{Ir} + \mathrm{O}_2 \leftrightarrow \mathrm{IrO}_2 \quad \Delta H = -240.2 \,\mathrm{kJ/mol} \tag{6}$$

where the remarkably negative $\Delta H$ shows that at room temperature iridium, although a member of the platinum group of metals, is thermodynamically unstable against oxygen.

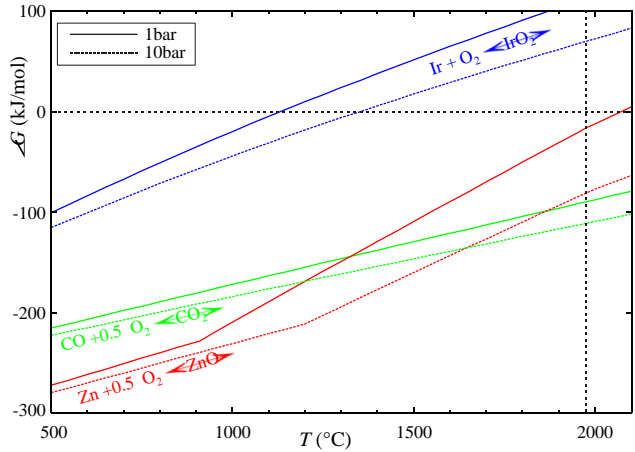

**Fig. 6: Gibbs free energy balance for the 3 reactions that are relevant for ZnO growth from Ir crucibles within reactive $CO_2$ atmosphere for 1 bar and 10 bar total pressure. The vertical dashed line marks the melting point of ZnO $T_f$ = 1975°C.**

The Gibbs free energy balance for ZnO (1) and $IrO_2$ (6) formation are plotted for two pressures in Fig. 6. It was already mentioned that the evaporation of ZnO under ambient pressure is so high that it sublimes without previous melting, but already under an overpressure of a few bar the ZnO melt can be kept stable. The following points are read from Fig. 6:

- Ir oxidation becomes unfavorable above ca. 1100-1300°C. Therefore for very high temperatures corrosion by oxygen will only be marginal.

- Zn oxidation is favored only for low $T$. The tendency of ZnO to dissociate to Zn + $O_2$ (see also Fig. 2) can be suppressed to a certain degree by application of overpressure.

- In the whole range where Ir could be oxidized ($T < 1300°C$) the dissociation of ZnO is almost completely prohibited because $CO_2$ dissociates more. This means that in a $CO_2$ atmosphere no liquid Zn can be formed (boiling point ca. 1200°C) that would alloy and destroy the crucible.

For growing crystals, pressed ZnO powder is placed inside an Ir crucible with conical bottom and seed channel, as shown in Fig. 5c. The crucible is surrounded by ceramic insulation (not shown in the figure) and is heated by an *rf* induction coil. If the crucible content (except the seed on the bottom) is molten, the heating power is lowered to initiate seeded growth. So far, single crystalline ZnO boules up to 38 mm diameter were grown [17].

### **From solutions, including hydrothermal**

So far, three types of solvents were reported for ZnO bulk growth:

1. Aqueous fluxes at ambient pressure are often based on highly concentrated solutions of KOH and/or other alkaline hydroxides where different zincates such as described by eq. (5) are formed. The handling of the very aggressive hydroxides is not easy, and besides KOH tends to absorb water and carbon dioxide from air. Resulting from such impurities, technical

KOH was found to melt at 140°C, much below the melting point of the pure substance of 243°C. Often silver crucibles are used because this metal is more stable than even platinum. During growth runs of 2 days at 450-480°C transparent ZnO needles with 6 mm length and 0.2 mm diameter could be grown [18]. By adding LiOH to the solvent also single crystalline plates (diameter 5 mm, thickness 0.1 mm) can be obtained [19]. Usually for all such experiments $\vec{c}$ is either the needle axis or the plate normal.

2. Typical melt fluxes composed e.g. of vanadates and phosphates [20], $MoO_3$, $B_2O_3$ [21], $PbF_2$ or PbO [22] were used for growing ZnO also. Unfortunately, flux grown crystals are always contaminated by solvent, sometimes up to 1% [21], which makes their application e.g. for electronic devices difficult. An interesting variant allows using a self-flux of molten zinc to grow ZnO crystals [23]. Only small crystals could be obtained, but the purity was reported to be higher compared with commercial hydrothermal material.

3. Hydrothermal solutions are chemically comparable with the alkaline aqueous fluxes that were mentioned in point 1, but have lower hydroxide concentration. Exact data on process parameters are usually undisclosed by the producers. Hydrothermal growth means conditions beyond the critical point of the solvent. For pure water this is $T_c$ = 373.9°C, $p_c$ = 220.6 bar. In such reactors a polycrystalline feedstock (powder or compacted grains) is heated together with the solvent inside an autoclave in such a way, that a small temperature gradient of ca. 10−15 K transports the nutrient to the growth zone where seeds are initiating growth. The process is similar to the technology for mass production of α-quartz crystals that was developed at Bell Laboratories after World War II [24] and is nowadays the almost exclusive source of commercial ZnO bulk crystals. Mainly groups in Japan and Russia grow crystals with several inch diameter in platinum or Ti-alloy lined autoclaves with inner diameter of ≥200 mm and length up to several meters [25,26]. The crystallographic quality of hydrothermally grown crystals is often impressive (etch pit density a few 100 $cm^{-2}$ or even lower, FWHM of (0002) as low as 8 arcsec were reported [25]). One should not forget, however, that hydrothermal growth is still a solution growth technology, with the typical incorporation of solvent traces (especially alkaline metals and hydrogen) in the grown crystals.

**RELATED COMPOUNDS: OXIDES OF GALLIUM, INDIUM, TIN**

Besides ZnO, oxides of some other metals (typically from the 3$^{rd}$ or 4$^{th}$ main group) are promising materials e.g. for **T**ransparent **C**onductive **O**xide applications [27]. Often these materials are used as polycrystalline layers, but at least for fundamental research single crystals are desirable. Fig. 7 compares stability fields of three relevant materials with that of ZnO. The topmost black dashed line shows $p_{O_2}(T)$ that is delivered by carbon dioxide at ambient pressure. For all metal-oxygen systems only the phase boundary separating the desired oxide from the corresponding metal that is formed for too low $p_{O_2}$ is shown. For the melt growth of the oxides the system must be kept stable near the melting point that is indicated by a short line towards the oxide side of the phase boundary.

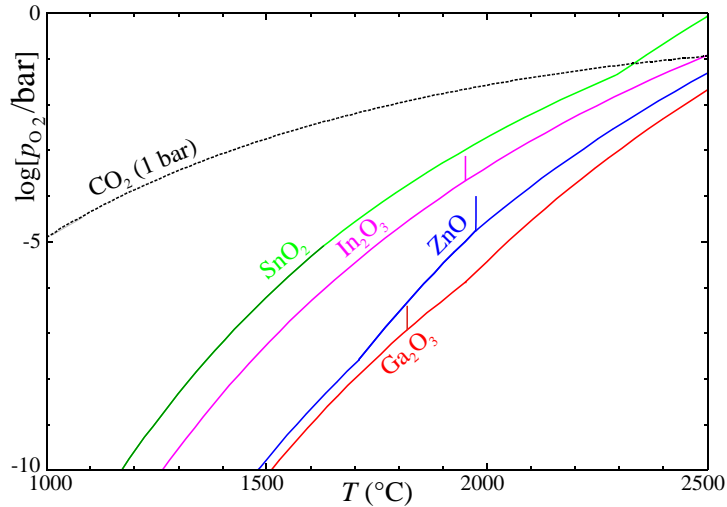

**Fig. 7:** Stability fields for the TCO's $SnO_2$, $In_2O_3$, ZnO, and $Ga_2O_3$ (from top to bottom) compared with the oxygen fugacity that is delivered by pure $CO_2$. The short vertical dashes denote the melting points of $In_2O_3$ [28], ZnO [15], and $Ga_2O_3$ [29]

It is obvious that the requirements for the growth of $Ga_2O_3$ crystals are somewhat softer, compared with ZnO: the melting point is lower and less oxygen is required to minimize or avoid decomposition. The feasibility of the Czochralski growth for $\beta$-$Ga_2O_3$ was demonstrated already a decade ago [30] and boules being suitable for the production of 10×10 mm² wafers were grown recently [29]. The requirements for $In_2O_3$ are much harder, because about one order of magnitude more oxygen is required to stabilize this oxide at the melting point, compared with ZnO. Nevertheless, the first truly bulk $In_2O_3$ single crystals grown from the melt were demonstrated [28]. Attempts to grow $SnO_2$ crystals from the melt failed so far.

Unfortunately for $SnO_2$ the current knowledge even about the melting temperature is insufficient. Often values around 1630°C are reported [31], but this was proven to be wrong by the current authors. Pure $SnO_2$ powder heated in flowing oxygen in a NETZSCH STA409 to 1650°C showed considerable evaporation but no melting. One can assume that partial oxygen loss of $SnO_2$, and subsequent melting of Sn, resulted in wrong (and too low) melting point data found in literature. Because the stability limit of $SnO_2$ (green curve in Fig. 7) approaches the oxygen partial pressure that is supplied even by pure $CO_2$ it remains questionable whether melt growth of pure tin dioxide from metallic crucibles will ever be possible.

"ITO" (indium tin oxide) is certainly one of the most important TCO's that is used in polycrystalline form e.g. for transparent electrodes in flat panel displays or solar cells. At room temperature this is a mixture of $In_2O_3$ saturated with Sn, and $SnO_2$ saturated with In. The solubility limit for both dopants is ≈10% [32]. Only for $T > 1200°C$ an intermediate compound with homogeneity range around $In_2SnO_5$ is stable. For lower $T$ this phase undergoes decomposition.

It would be desirable to identify systems where such intermediate compounds, or at least solid solutions, are stable or metastable down to room temperature. Already in the case of ideal miscibility the free enthalpy of the solid (also of the liquid) phase decreases then by

$$\Delta G_{\text{mix}} = -RT(x_1 \ln x_1 + x_2 \ln x_2)$$
$$\approx -0.7RT \text{ (at } x_1 = x_2 = 0.5\text{)}$$
(7)

The oxygen partial pressure in the system corresponds to a Gibbs free energy

$$\Delta G = -RT \ln p_{O_2} \qquad (8)$$

This mean that under the assumption of ideal mixing the oxygen partial pressure that is necessary to stabilize the condensed phases is lowered by $\ln[p_{O_2}] \approx 0.7$, and hence $\log[p_{O_2}] \approx 0.3$. With other words, the stability limit of an ideal mixture of two oxides such as shown in Fig. 7 would be shifted by ca. 1/3 tick mark downwards. Thus even in cases where pure oxides are not accessible, mixture phases could perhaps be grown.

## CONCLUSIONS

Bulk zinc oxide single crystals can be grown by many crystal growth techniques, but nowadays the hydrothermal technique with growth from alkaline aqueous solutions is mostly utilized. The crystals have up to several inch diameter, combined with good crystallographic quality. It should not be forgotten, however, that this is a solution growth method and that traces of the solvent (alkalines, hydrogen) are always incorporated into the crystal. Especially for hydrogen the severe influence on the electronic properties is widely discussed, and hydrogen seems to be one of the reasons why *p*-type doping of ZnO is so difficult.

It should be kept in mind that nearly all other semiconductor crystals with technical relevance (e.g. Si, GaAs) are grown from melts, and so melt grown ZnO could be an alternative for the standard hydrothermal material. Besides ZnO, the oxides of gallium, indium and tin are prospective TCO materials. At least $Ga_2O_3$ and $In_2O_3$ can now be grown from the melt, whereas $SnO_2$ remains a challenge.

## ACKNOWLEDGMENTS

The authors acknowledge financial support from the European Community under project ENSEMBLE NMP4-13SL-2008-213669.